\documentclass[aps,prl,twocolumn,showpacs,superscriptaddress,
groupedaddress]{revtex4-1}  % for review and submission

\usepackage{graphicx}  % needed for figures
\usepackage{amssymb}   % for math
\usepackage{natbib}    % references handling
\usepackage{color}     % colored text

\begin{document}

\newcommand{\ve}[1]{\mathbf{#1}}
\newcommand{\ket}[1]{\left| #1 \right\rangle}
\newcommand{\bra}[1]{\left\langle #1 \right|}
\newcommand{\braket}[3]{\left\langle #1 \right| #2 \left| #3 \right\rangle}

\title{Controlled energy-selected electron capture and release in double 
quantum dots}
\author{Federico M. Pont}
\email{federico.pont@pci.uni-heidelberg.de}
\author{Annika Bande}
\email{annika.bande@pci.uni-heidelberg.de}
\author{Lorenz S. Cederbaum}
\email{lorenz.cederbaum@pci.uni-heidelberg.de}
\affiliation{Theoretische Chemie, Physikalisch-Chemisches Institut,
Im Neuenheimer Feld 229, 69120 Heidelberg, Germany}
\vskip 0.25cm
      
\date{\today}

\begin{abstract}
Highly accurate quantum electron dynamics calculations demonstrate that energy 
can be efficiently transferred between quantum dots. Specifically, in a
double quantum dot an incoming electron is captured by one dot and the
excess energy is transferred to the neighboring dot and used to remove
an electron from this dot. This process is due to long-range electron
correlation and shown to be operative at rather large distances between the
dots. The efficiency of the process is greatly enhanced by preparing the
double quantum dot such that the incoming electron is initially captured by
a two-electron resonance state of the system. In contrast to atoms and molecules
in nature, double quantum dots can be manipulated to achieve this
enhancement. This mechanism leads to a surprisingly narrow distribution of the
energy of the electron removed in the process which is explained by
resonance theory. We argue that the process could be exploited in practice.
\end{abstract}

\pacs{73.21.La, 73.63.Kv, 34.80.Gs, 31.70.Hq}
\maketitle

Quantum dots (QDs) are often referred to as artificial 
atoms~\cite{kastner_artificial_1993} and double quantum dots (DQDs) are the
extension to molecules of this analogy~\cite{van_der_wiel_electron_2002}.
The initial notion for the name came from the quantized levels and transitions 
of carriers inside nanosized semiconductor 
structures~\cite{reed_observation_1988} that resemble those found
in atoms. A wealth of other phenomena also present in atoms have found their
counterpart in QDs~\cite{van_der_wiel_electron_2002, salfi_electronic_2010,
laird_coherent_2010, roddaro_manipulation_2011, nadj-perge_spectroscopy_2012}
together with new phenomena handed down from the semiconductor nature of
QDs~\cite{fujisawa_spontaneous_1998,
shabaev_multiexciton_2006, muller_electrical_2012, benyoucef_single-photon_2012}
many of which endowed new technological applications to be cast into
reality. In this work we concentrate on energy transfer between two QDs driven
by long-range electron correlation and mediated by the capture of an
electron.

Electron capture in single QDs is an extensively studied
topic~\cite{nozik_semiconductor_2010, prasankumar_ultrafast_2009, 
porte_ultrafast_2009} due to its relevance in the development of technological applications.
The capture efficiency and its time scale depend substantially
on temperature, carrier density, material and geometry
of the QDs~\cite{prasankumar_ultrafast_2009, porte_ultrafast_2009,
narvaez_carrier_2006}. The capture and the later relaxation
dynamics occur through diverse physical processes such
as electron-phonon interactions~\cite{seebeck_polarons_2005, prasankumar_ultrafast_2009,
porte_ultrafast_2009, zibik_long_2009}, multiple exciton
generation~\cite{nozik_semiconductor_2010} and Auger
relaxation~\cite{narvaez_carrier_2006}, all of
which can be assessed using pump-probe
schemes~\cite{prasankumar_ultrafast_2009,porte_ultrafast_2009,
zibik_long_2009,nozik_semiconductor_2010,muller_electrical_2012}. 
Capture by
optical phonon emission has been investigated in
single~\cite{glanemann_transport_2005, seebeck_polarons_2005, jiang_inelastic_2012} as well as
in double QDs~\cite{glanemann_transport_2005}.
So far, electronically-induced
inter-dot capture processes have not been considered at all. 
In the present work we use numerically exact quantum dynamics to show that electron
capture by one QD in a DQD becomes possible by energy
transfer to the neighboring QD due to long-range electron
correlation. Originally, such processes were predicted to
operate between atoms~\cite{gokhberg_environment_2009,
gokhberg_interatomic_2010} where electron capture
by one atom occurs while another electron is emitted
from an atom in its environment and called interatomic
electronic Coulombic capture (ICEC), a name which we
would like to adopt also for QDs. For completeness we mention that
energy transfer between quantum wells has been
studied in a different context, see,
\emph{e.g.},~\cite{batsch_dipole-dipole_1993, tomita_efficient_1996}.

\begin{figure}
\includegraphics[width=0.45\textwidth]{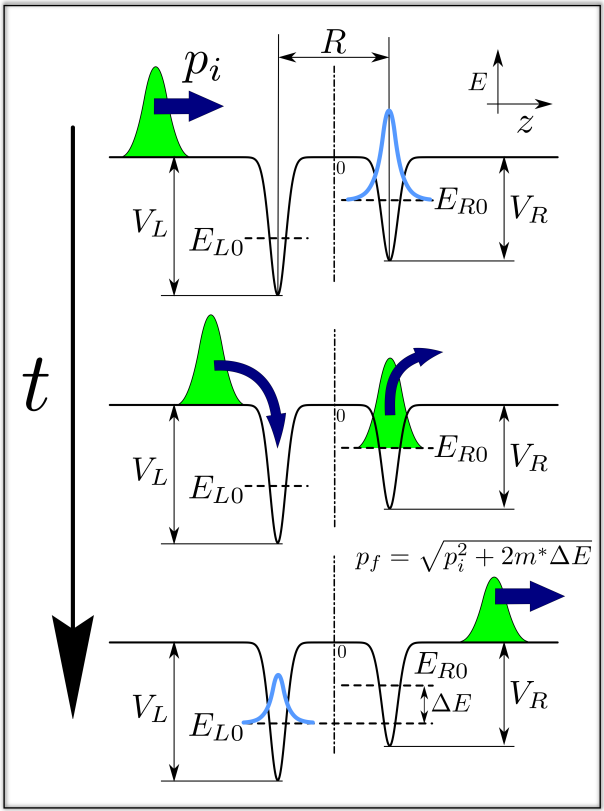}
\caption{\label{scheme}(Color online) Schematic view of ICEC in a model 
potential for a DQD. The capture of the incoming electron by the left QD is
mediated by its correlation with the electron initially bound to the right QD.
After energy transfer, the electron on the right is excited into the continuum
and becomes an outgoing electron.}
\end{figure}

For explicit demonstration we study a system comprised of two different QDs
which we call the left and right QDs and which are described by the model
potentials discussed below. Let the left potential well support a one-electron 
level $L_0$ and the right one, $R_0$. Although
included in the calculation, the tunneling between $L_0$ and
$R_0$ is vanishingly small due to the long inter-dot distance.
As described in Fig.~\ref{scheme}, an electron is initially in the right
QD and an electron with momentum $p_i$ is incoming from
the left of the DQD. This electron is captured into the $L_0$ ground state
of the left QD while the excess energy is transferred to the right QD emitting
the electron from the $R_0$ ground state of this QD. According to
energy conservation~\cite{gokhberg_interatomic_2010}
\begin{equation}\label{energy_cons}
E_{R_0} + \varepsilon_{i} = E_{L_0} + \varepsilon_{f}
\end{equation}

\noindent the momentum of the outgoing electron is 
$p_f=\sqrt{p^2_i+2m^*(E_{R_0}-E_{L_0})}$ where 
$\varepsilon_{{i,f}}=p^2_{i,f}/2 m^*$ and $m^*$ is the electron effective mass
in atomic units. The emitted electron can have a higher or lower momentum than
the initial electron, depending on the relation between the bound-state energies
$E_{R_0}$ and $E_{L_0}$.

Below, we will first discuss the model potentials of the two QDs and then prove
numerically that ICEC takes place. The findings will actually make clear that the 
process is, in principle, possible for two-site systems with a broad range
of binding potentials. Subsequently we investigate how
to manipulate DQDs in order to make the probability of ICEC large. This will
lead to a slightly more complex physical situation with an interesting energy
transfer.

The model potential for the DQDs is based on the effective
mass approximation~\cite{bastard_wave_1991} and allows to describe
accurately the motion of two electrons inside the nanostructured semiconductor.
Thus, it offers straightforward observability of how electron correlation
can lead to ICEC. In the specific DQD model adopted from a previous
study~\cite{bande_dynamics_2011} the dots are represented by two Gaussian wells
aligned in $z$ direction. In $x$ and $y$ direction we assume a strong harmonic
confinement which could be attributed either to depleting
gates~\cite{fujisawa_spontaneous_1998} or to the structure of the
semiconductor~\cite{salfi_electronic_2010}. The system
Hamiltonian is
\begin{equation}\label{ham}
H(\ve{r}_1, \ve{r}_2) = h(\ve{r}_1) + h(\ve{r}_2) +
\frac{1}{\left|\ve{r}_1-\ve{r}_2\right|}
\end{equation}

\noindent where
\begin{equation}\label{1eham}
h(\ve{r}_i) = -\frac{1}{2 m^*}\nabla_i^2 + V_{c}(x_i,y_i) +
V_{l}(z_i)
\end{equation}

\noindent is a one-electron Hamiltonian and
\begin{eqnarray}
V_{c}(x_i,y_i) &=& \frac{1}{2}m^*\omega^2(x_i + y_i)^2 \\
V_{l}(z_i) &=& - V_L e^{-b_L (z_i + R/2)^2} - V_R e^{-b_R (z_i - R/2)^2}
\end{eqnarray}

\noindent are the transversal confinement and longitudinal open potentials,
respectively. $R$ is the distance between the QDs, $b_{L,R}$ the sizes of
the left and right QDs and $V_{L,R}$ are their depths. Due to the strong
confinement ($\omega=1.0$ a.u. $> V_{L,R}$) the relevant excitations are 
only in $z$ direction into the levels $L_n(R_n)$, $n=0,1, \ldots$ of
the left (right) QD with energies $E_{L_n}(E_{R_n})$. For simplicity, we limit
ourselves here to triplet configurations, but mention that we do have
calculations on singlet configurations which show similar behavior.

The dynamical evolution of the system was obtained by solving the time-dependent
Schr\"odinger equation employing the multiconfiguration time-dependent Hartree
(MCTDH) approach~\cite{meyer_multidimensional_2009,
meyer_multi-configurational_1990}. The triplet wave function
\begin{equation}
\Psi(\ve{r}_1,\ve{r}_2,t) = \sum_{i,j} A_{ij}(t) \phi_{i}(\ve{r}_1,t)
\phi_{j}(\ve{r}_2,t),
\end{equation}

\noindent was expanded in time-dependent single particle functions
$\phi_{i}(\ve{r},t)$~(SPFs) and coefficients $A_{ij}(t)$ that fulfill the
antisymmetry condition $A_{ij}(t)=-A_{ji}(t)$. The SPFs $\phi_i$
were expanded in one-dimensional basis functions for each of the Cartesian
coordinates $(x,y,z)$ in a DVR-grid representation (discrete variable
representation) with harmonic oscillator DVRs for $x$ and $y$, and a sine DVR
for the $z$ coordinate. The MCTDH equations of motion for coefficients and SPFs
were efficiently solved using the approach implemented in the MCTDH-Heidelberg
package~\cite{meyer_multidimensional_2009, beck_multiconfiguration_2000}. A
complex absorbing potential (CAP) of order~$2$ was placed in the $z$ coordinate
to absorb the outgoing electron before it reaches the end of the
grid. We placed CAPs at different positions and confirmed that the
results discussed below are not affected by the CAP. The
Coulomb potential was regularized  to prevent divergences at
$\ve{r}_1=\ve{r}_2$, $1/r_{12} \rightarrow 1/\sqrt{r^2_{12}+a^2}$ with $a=0.01$,
and then transformed into sums of products using the
POTFIT~\cite{beck_multiconfiguration_2000} algorithm of MCTDH. The convergence
of numerical results was ensured.

\begin{figure}
\includegraphics[width=0.48\textwidth]{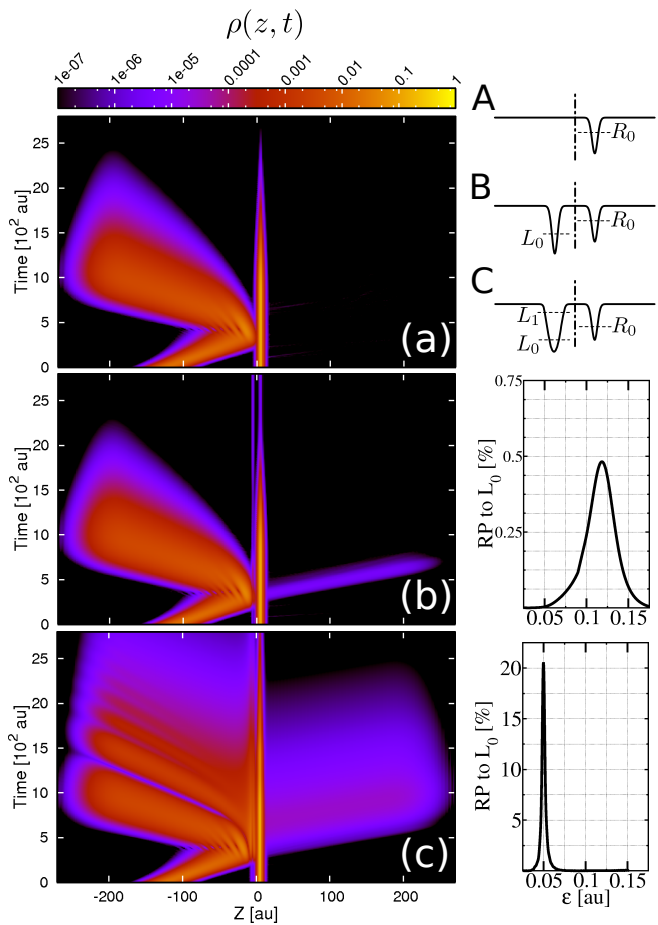}
\caption{\label{surface_plots} (Color online) 
Evolution of the electronic density (left panels)
and reaction probabilities (RP) (right panels) for three different setups shown
in the upper right corner. The incoming wave packet (WP$_i$) approaches from the
left to the DQD centered at $z=0$~a.u. which is initially occupied by
an electron in the right QD ($R_0$ state).
Both WP$_i$ and $R_0$ are the same in all three cases. The complex
absorbing potentials (CAPs) are placed at $z=\pm 170$~a.u.
(a) The left QD is missing and no emission to the right is observed.
(b) The left QD has a single one-electron state $L_0$.  Emission of the electron
initially in $R_0$ through ICEC takes place. The emitted packet acquires
momentum according to the energy conservation, Eq.~(\ref{energy_cons}), and is 
faster than the incoming packet in this case.
(c) The left QD supports two one-electron states and the DQD exhibits a
two-electron $\ket{L_1R_0}$ resonance state which is populated and decays,
strongly enhancing the RP of the ICEC channel. The RP shows a Breit-Wigner
shape.}
\end{figure}

Let us discuss the results. We first investigate what
happens if only the right QD is present. The result is depicted in
Fig.~\ref{surface_plots}(a) (setup $A$). The parameters $V_R=0.6$~a.u. and
$b_R=1.0$~a.u. used  give a single bound state with an energy of
$E_{R_0}=-0.246$ a.u. (we set the origin of the energy scale to $2\hbar\omega=2$
a.u. throughout the study). The incoming wave packet (WP$_{i}$) was represented
by an energy normalized Gaussian peaked around $\varepsilon_{WP_i}=0.056$ a.u.
with energy width $\Delta E_{WP_i}\approx 0.033$~a.u. and spatial width
$\Delta x_{WP_i}=10.0$~a.u. The ionization of the bound electron by the 
incoming one and the excitation to higher states in the transversal directions 
are energetically forbidden for these parameters.

The dynamics of the scattering process is visualized
in Fig.~\ref{surface_plots}(a) by the longitudinal electronic density
$\rho(z,t)=\int \textrm{d}\ve{r'} \int\textrm{d}x \int\textrm{d}y
\left|\Psi(\ve{r},\ve{r'},t)\right|^2$ as a function of $z$ and
$t$. It is clearly seen that the incoming electron is completely reflected
while the other electron remains bound in the right QD.

We now add the left QD at a distance $R = 10.0$~a.u. and show that ICEC takes
place in the DQD according to the scheme in Fig.~\ref{scheme}. We choose 
the potential well of the left QD somewhat deeper ($V_L=0.8$ a.u.,
$b_{L,R} = 1.0$~a.u.) than that of the right QD while leaving the parameters of
the latter unchanged (setup $B$). This choice leads to two non-overlapping bound
one electron states, $L_0$ and $R_0$, with energies                          
$E_{L_{0}}=-0.377$~a.u. and $E_{R_{0}}=-0.246$~a.u.

The spatially resolved time evolution of $\rho(z,t)$ in
Fig.~\ref{surface_plots}(b) clearly shows emission of electronic density to the
right of the DQD starting at $t\approx 400$~a.u. 
The slope of the final wave packet (WP$_{f}$)  trajectory traveling to the right changed
compared to that of the incoming WP$_i$ indicating that the emitted electron has
acquired momentum during the process.
%The slope of the final wave packet (WP$_{f}$)  trajectory traveling to the right indicates 
%that the emitted electron has higher momentum than that of the incoming WP$_i$.
From the slope we estimate an
average momentum $p^{(plot)}_{f}\approx 0.63\pm0.02$~a.u. which is in full
agreement with energy conservation in ICEC (see Eq.~(\ref{energy_cons})),
where $\Delta E =E_{R_0}-E_{L_0}= 0.131$~a.u. gives $p_f=0.612$~a.u. We also
computed examples for $\Delta E < 0$ that showed, accordingly, the emission of
decelerated electrons to the right. 
    
Obviously, ICEC takes place in the DQD. To have a quantitative measure, we
computed the reaction probability (RP). The RP is a function of the energy of
the incoming electron and tells us the probability with which an incoming
electron with a given energy is caught in the left well and simultaneously an
electron is emitted to the right from the right well. Of course, the energy of
the emitted electron is regulated by Eq.~(\ref{energy_cons}). Technically this
is done by computing the energy-dependent absorbed flux into the right CAP after
normalization with the distribution of the incoming wave packet
WP$_i$~\cite{beck_multiconfiguration_2000, meyer_multidimensional_2009}. Due to
this normalization, the RP is \emph{independent} of WP$_i$ which makes it an
absolute measure. The RP corresponding to $\rho(z,t)$ depicted in
Fig.~\ref{surface_plots}(b) is shown at the r.h.s. of the panel. It peaks at an
energy of about ${\varepsilon_i}=0.12$~a.u. where it acquires a value of
$0.5\%$. It is seen that ICEC in QDs is selective in energy. The peak in the RP
appears close to the total energy $E_T= \varepsilon_i + E_{R_0}$ which fits to
the energy transferred to the QD system by ICEC: $E^{peak}_T=E_{L_0}-E_{R_0}$.
Considering that the capture of a freely moving electron is a rather 
intricate process in 
general~\cite{prasankumar_ultrafast_2009, porte_ultrafast_2009,
sauvage_long_2002}, a probability of $0.5\%$ is rather high. On the other hand,
it is still a rather low value if ICEC is to be utilized in practical
applications.

How to enhance the reaction probability of ICEC substantially ? To answer this
relevant question we make use of the physics of resonance states. To that end we
widen the potential well of the left QD by choosing $V_L=0.71$ and
$b_L=0.3$~a.u. such that this dot displays a bound excited one-electron state
$L_1$ in addition to its ground state $L_0$ (see upper right corner of
Fig.~\ref{surface_plots} (setup $C$)). The energies in this case
are $E_{L_{1}}=-0.048$ and $E_{L_{0}}=-0.441$~a.u. The incoming electron
properties were kept unchanged and we notice that the energy of WP$_i$ is
insufficient to open the ICEC channel to the $L_1$ state even if we take into
account the energy width of this wave packet. The ICEC channel
to the $L_0$ state is, of course, still open. However, due to the lower energy
of $L_0$, the RP of ICEC directly populating that level is even lower than that
in case $B$. 

The results of the propagation in setup $C$ are shown in
Fig.~\ref{surface_plots}(c). It is eye catching that the electron density 
emitted to the right is now much larger than in case $B$. The RP to the 
right (r.h.s. of Fig.~\ref{surface_plots}(c)) is now amplified and
peaks at $20\%$ at the energy $\varepsilon_{i}=0.05$~a.u. of the incoming
electron. Compared to case $B$, this is an amplification by a factor of 40! We
remind that the RP is independent of the incoming wave packet used. There is
another very interesting property of the RP which can be relevant for practical
applications, namely that the RP is very narrow in energy, much narrower than in
case B. 

Let us briefly discuss the density plot in Fig.~\ref{surface_plots}(c) which 
also shows several unusual features. From the plot the momentum gives
$p^{(plot)}_f \approx 0.70\pm 0.02$~a.u. in agreement with $p_f=0.70$~a.u., which
follows from Eq.~(\ref{energy_cons}). In sharp contrast to case $B$, the WP$_f$
is not created in a relatively sharp instant of time (compare
Figs.~\ref{surface_plots}(b) and \ref{surface_plots}(c) at the site of the DQD, 
$z = 0$) but rather continuously with an exponential decay in time which gives a
hint about the mechanism of amplification which will be discussed below. The
emitted electron density to the left side becomes more complex compared to
Fig.~\ref{surface_plots}(b) and shows signatures of interference which we
attribute to the superposition between the reflection of the incoming wave
packet WP$_i$ and electrons emitted to the left in the same energy range.

The substantial amplification of electron capture by the left QD and the other
features mentioned above can be well understood by realizing the interplay of
two effects. First, the appearance of a resonant state and second that this
state efficiently decays to the ICEC channel. 

Due to the additional one-electron level $L_1$, the DQD of setup $C$
accommodates a \emph{two-electron} resonance $\ket{L_1R_0}$, which has one
electron on each of the QDs. Such states have been shown to decay fast by energy
transfer to $\ket{L_0}$ plus an outgoing electron via interatomic Coulombic
decay (ICD)~\cite{bande_dynamics_2011, cherkes_electron_2011,
cederbaum_giant_1997, sisourat_interatomic_2010, sisourat_ultralong-range_2010,
jahnke_ultrafast_2010}. The
resulting story is then 
that if the incoming electron is in the energy range of the two-electron resonance state, 
it has a higher probability to be caught in this state than
capturing of one electron and emitting another electron. This state can decay by
ICD thus strongly amplifying the RP of the ICEC channel.

Let us discuss how this picture explains our observations in
Fig.~\ref{surface_plots}(c). The energy of the $\ket{L_1R_0}$ can be estimated
for large inter-dot distances $R$ as $E_{L_1R_0}\approx E_{R_0}+E_{L_1} + 1/R$. 
In order to be substantially populated by the scattering process,
this two-electron energy should correspond to the total energy $E_T=\varepsilon_{i}+E_{R_0}$. 
For the parameters used, $E_{L_1R_0}\approx 0.052+E_{R_0}$~a.u., which is indeed very close to
the peak of the RP which is at $\varepsilon_i=0.05$~a.u. The exponential decay in time observed above for
the continuous emission of electrons to the right can be attributed to the
lifetime of the $\ket{L_1R_0}$ resonance populated by the incoming electron. 

Decay rates (inverse lifetimes) in QDs can be computed using different
methods~\cite{bande_dynamics_2011, cherkes_electron_2011, pont_entropy_2010}.
Here, we employed imaginary time propagation to arrive at the resonant state of setup  $C$ and then
the state is let to evolve in real time to measure the total decay 
rate~\cite{bande_dynamics_2011}. The
decay rate obtained for case $C$ is $\Gamma_{C}=39\pm2 \times 10^{-4}$~a.u. For
consistency we have also fitted a Breit-Wigner peak
shape~\cite{taylor_scattering_2006} to the RP in
the r.h.s. of Fig.~\ref{surface_plots}(c) and obtained 
$\Gamma^{(RP)}_{C}=38\pm1 \times 10^{-4}$~a.u. which is in perfect agreement
with the results from the propagations. Finally, we mention for completeness
that the resonance after being populated by the incoming electron can also decay
by emitting elastically the electron to the left resembling that of a shape
resonance~\cite{taylor_scattering_2006}: 
$e^{-} + \ket{R_0}\rightarrow \ket{L_1R_0} \rightarrow \ket{R_0} + e^-$. This
electron is responsible for the interference effects mentioned above. 
Our calculations show, however, that this
depopulation channel is minor in comparison to the ICEC channel.

Having proven that ICEC takes place, we now transfer the parameters to
realistic semiconductor QDs. The quasi one-dimensional shape used in the model
is applicable to experimentally achievable DQDs, for example QDs embedded in
nanowires~\cite{salfi_electronic_2010} or electrostatically defined
dots~\cite{fujisawa_spontaneous_1998}. The process is driven by
long-range 
Coulomb interactions, so we expect ICEC to be also applicable to other QDs
geometries like, \emph{e.g.}, self-assembled vertically stacked
dots~\cite{muller_electrical_2012, benyoucef_single-photon_2012,
porte_ultrafast_2009, zibik_long_2009}. We convert the atomic units of setups $B$ and $C$
into units of GaAs QDs using the effective mass approximation~\cite{bande_dynamics_2011}. 
Then
$R^{GaAs}\approx 98$~nm, 
$\Delta E^{GaAs}(B) = 1.55$~meV, 
$\Delta E^{GaAs}(C) = 2.30$~meV and 
$\Delta E^{GaAs}_{L_1R_0}(C)=0.61$~meV and for 
WP$_i$ we have $\varepsilon_{WP_i}=0.66$~meV and 
$\Delta E_{WP_i}\approx 0.4$~meV. 
These energies are well in the range of intraband level spacings of QDs
in nanowires~\cite{salfi_electronic_2010,roddaro_manipulation_2011} and of 
intrashell levels in self-assembled QDs~\cite{zibik_long_2009}.
The time scale depicted in Fig.~\ref{surface_plots} is $160$ ps for GaAs QDs.
As seen in the figure,  ICEC emission occurs for case B on a surprisingly short
time scale of 10 ps. This is much faster than the reported capture
times of 100 ps for free carriers in bulk GaAs into InAs/GaAs QDs in single
layer samples measured at room temperature~\cite{turchinovich_inas/gaas_2003}. 
Notice that electron capture by emission of optical phonons
is ineffective in our case where the phonon energy is much larger than the
electronic transition~\cite{glanemann_transport_2005, seebeck_polarons_2005}.
The lifetime of the $\ket{L_1R_0}$ resonance in case 
$C$ is $14.3$ ps and thus also short. 
This is important because the decay of this resonance may compete
with relaxation via phonons. The times for ICEC are, however, faster than
reported intraband decay times due to acoustic phonon emission for InGaAs/GaAs
QDs~\cite{zibik_long_2009}. It is relevant to mention that the
width of the RP peak in Fig.~\ref{surface_plots}(c) is only $0.046$~meV.

In summary, fully correlated electron dynamics was
used to show that long distance energy transfer between
the quantum dots of a DQD is possible due to long-range
electron correlation. The transfer is achieved by a fundamental
electronically-induced process where capture of
an electron in one QD induces a release of another electron from
a distant quantum dot. This fundamental process turns
out to be particularly fast and can overcome
other important capture mechanisms such as acoustic
phonon emission. The presence of a two-electron resonance
in the DQD results in a substantial enhancement of the energy transfer
and leads to a well defined and narrow energy distribution of the
emitted electron. The ICEC mechanisms in DQDs are not only interesting from the point of view of basic
physics, but could, in principle, also be exploited to
implement devices which generate nearly monochromatic
low energy electrons in a given direction. We think that
the implementation can be based on currently available
nanowires, particularly those with long free electron
lifetimes.

We thank K. Gokhberg for discussions and H.-D. Meyer for help with MCTDH. 
A. B. acknowledges financial support by the Heidelberg University
(Olympia-Morata fellowship) and F. M. P. by Deutscher Akademischer
Austauschdienst (DAAD).

\bibliographystyle{apsrev4-1}
%\bibliography{letter.bib}

%

\end{document}